\documentclass[info]{epl}
\usepackage{amssymb,amsmath}

\title{Quantum impurity approach to a coupled qubit problem}
\author{S. Camalet\inst{1} \and J. Schriefl\inst{1,2} \and P. Degiovanni\inst{1} \and F. Delduc\inst{1}}
\institute{
  \inst{1} CNRS-Laboratoire de Physique, Ecole Normale Sup\'erieure de lyon - 46 all\'ee d'Italie, 69007 Lyon, France\\
  \inst{2} Institut f{\"u}r Theoretische
Festk{\"o}rperphysik - Universit{\"a}t Karlsruhe,
 76128 Karlsruhe, Germany }
\shortauthor{S. Camalet \etal}
\begin{document}

\maketitle

\begin{abstract}
We consider a system of two qubits at the ends of a finite
length 1D cavity. This problem is mapped onto the double-Kondo model which
is also shown to describe the low energy physics of a finite length quantum wire with
resonant levels at its ends. At the Toulouse point the ground state energy and the average populations
and correlations of the qubits or resonant
levels at zero temperature are computed. These results show that the
effective interactions between the qubits or resonant levels can be used to probe their associated
Kondo length scale.
\end{abstract}


Cavity quantum electrodynamics (cQED) concerns the study of atoms
coupled to discrete quantized electromagnetic modes in a high-$Q$
cavity. Recent experiments performed with Rydberg atoms in a
microwave cavity have shown how cQED can be used to create and
manipulate entanglement between atoms and the quantized
electromagnetic field (see \cite{Haroche:review} for a review).
Experiments performed on a variety of Josephson devices have demonstrated
their potentialities as artificial two-level systems suitable for
quantum state engineering (see \cite{Makhlin:2001-1} for a
review). These considerations motivated a recent proposal
\cite{Blais:2004-1} for a cQED-type quantum computing architecture
based on Cooper pair boxes embedded in a super-conducting
transmission line.

All these cQED schemes consider the coupling of a single quantum
electromagnetic mode to the qubits and the theoretical analysis is
performed using the Jaynes-Cummings Hamiltonian which relies on
the rotating wave approximation. However, in a sufficiently large
cavity a large number of modes exist.
Whereas in the weak coupling regime the single-mode
approach and the rotating wave approximation are expected to give
correct results, this is not the case when the qubits are strongly
coupled to the transmission line. For instance, Leclair
\etal~showed \cite{Leclair:1997-1,Leclair:1999-1} that the
exact spectrum of a single qubit coupled to an infinite
transmission line is considerably richer than the
spectrum of the Jaynes-Cummings Hamiltonian.
In the same model the average population and fluctuations of the qubit at zero
temperature have also been studied \cite{Cedraschi:2001-1}.
But so far, very little is
known on the correlations between several qubits strongly coupled
to such an environment. In this regime, neither the
Jaynes-Cummings Hammiltonian nor perturbation theory in the
qubit/cavity coupling are expected to work. The purpose of this
letter is to study these correlations in the strong coupling regime
using techniques and concepts developed in the context of quantum
impurity problems.

\begin{figure}
\twofigures[scale=0.25]{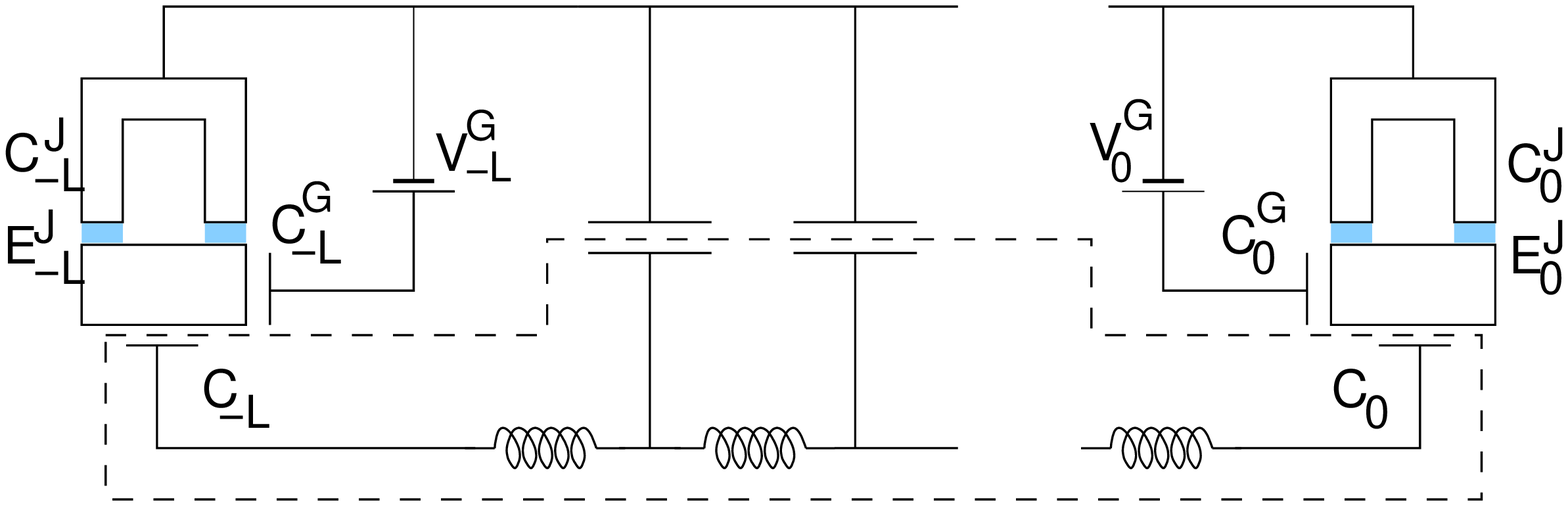}{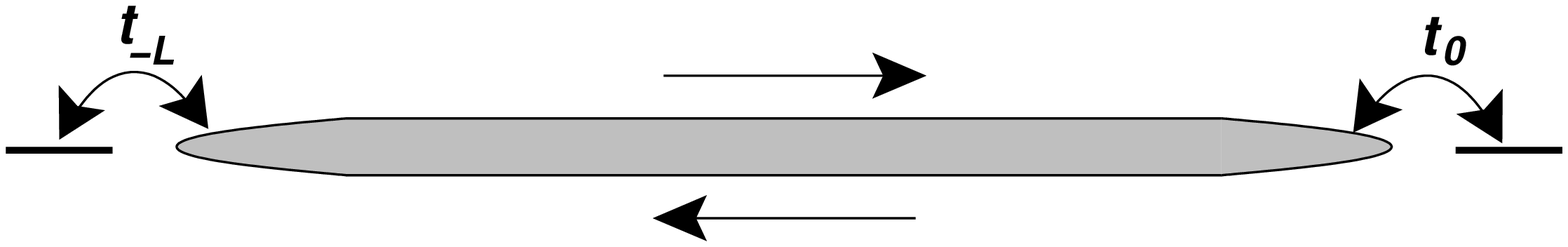}
\caption{Two Cooper pair
boxes capacitively coupled to a quantum transmission line (equivalent circuit).
The electric charge stored in the
dashed box is conserved. \label{fig.1}}
\caption{Two resonant levels coupled by tunnel junctions to the ends of a
quantum wire of length $L$.
\label{fig.2}}
\end{figure}

In this letter, we consider the coupling of two qubits through a
transmission line of length $L$ (see fig. \ref{fig.1}). Using a
mapping onto the double-Kondo model we study the correlations
between the two qubits induced by their interaction via the
transmission line.
At the Toulouse point of the double-Kondo model we obtain
an explicit description of the ground state and low energy
excitations of the system. The inter-qubit correlations at $T=0$~K
are computed exactly and shown to depend on
intrinsic length scales associated with each qubit
(Kondo lengths). Depending on the ratio of these Kondo lengths to
$L$ the system flows from a Kondo regime where each qubit renormalizes to a conformally
invariant boundary condition and is not influenced
by the other to a correlated regime where the degrees of freedom of the qubits get
dressed and coupled by the transmission line. This crossover occurs when
the Kondo lengths associated with each qubits are comparable to the
size $L$ of the system. Kondo cloud physics in a finite system
with a single quantum impurity has recently received attention \cite{Affleck:2001-1}. We point
out that any system described by the double-Kondo model provides a way to probe the interplay
between Kondo cloud physics and interactions
in a finite size system with two quantum impurities.

Originally introduced
in string theory within the context of tachyon instabilities
\cite{Bardakci:2001-1} the double-Kondo model not only describes the two-qubit problem
but may also be relevant in mesoscopic physics. We show that it describes the
low energy behavior of a finite length 1D quantum wire
coupled by tunnel junctions to two resonant levels (see fig.
\ref{fig.2}). The quantities of interest are the average
occupation numbers of each resonant level and their correlations.
The device of fig. \ref{fig.2} may be realized using conventional
lithography techniques. We think that it could provide a
complementary route to explore the physics of coupled quantum
impurities underlying the system of fig. \ref{fig.1}.

The quantum circuit of fig. \ref{fig.1} consists of two Josephson charge qubits
built from Cooper pair boxes in the charge regime
\cite{Makhlin:2001-1} capacitively coupled to a quantum
transmission line.
For convenience, each qubit is labelled by its position
along the transmission line: $j=-L$ or $0$.
In the continuum limit the quantum transmission line is described in
terms of a free bosonic field
$\Phi(x)$ and its conjugated field $\Pi(x)$ ($-L\leq x\leq 0$)
related to the charge and current densities along
the line:
\begin{equation}
\label{eq.1}
i(x,t) =
-ev\,\sqrt{\frac{\hbar/e^2}{\mathcal{R}}}\,(\partial_x\Phi)(x,t)
\quad\mathrm{and}\  \rho(x,t) =
\frac{e}{\hbar}\,\sqrt{\frac{\hbar/e^2}{\mathcal{R}}}\,\Pi(x,t)
\end{equation}
where $v=1/\sqrt{lc}$ is the wave velocity along the transmission
line and $\mathcal{R}=\sqrt{l/c}$ is the resistance of the
semi-infinite line both expressed in terms of the line's
capacitance $c$ and inductance $l$ per unit of length.
The low-energy effective Hamiltonian of the
system depicted on fig. \ref{fig.1} is then:
\begin{equation}
\label{eq.2} H = -\frac{1}{2}
\sum_{j=-L,0}(E_j^J\sigma_j^x+\tilde{B}_j^z\sigma_j^z) +
\frac{\hbar v}{2}
\int_{-L}^0\left(\hbar^{-2}\Pi^2+(\partial_x\Phi)^2\right)\upd x
+\frac{v}{2} \sum_{j=-L,0}\beta_j\Pi(j)\ldotp(\sigma_j^z+2n_j^G-1)
\end{equation}
where the first term represents the bare qubit Hamiltonian \cite{Makhlin:2001-1}.
The effective magnetic field $\tilde{B}^z_j=2e^2(1-2n^G_j)/(C^J_j+C^G_j)$
depends on the externally controlled
gate voltage $V_j^G$ ($n_j^G=C^G_jV_j^G/2e$). Here $C_j^J$ and
$C_j^G$ respectively
denote the Josephson junction and gate capacitances of qubit $j$.
The Josephson energy $E^J_j$ can
be controlled by applying an external flux through the loop of the
SQUID device.
The second term in (\ref{eq.2}) is
the free transmission line Hamiltonian. The third one couples the qubits to the transmission line with
dimensionless coupling constants
\begin{equation}
\label{eq.3}
\beta_j=\frac{2\,C_j}{C_j+C^J_j+C_j^G}\,\sqrt{\frac{\mathcal{R}}{\hbar/e^2}}\,.
\end{equation}
This Hamiltonian is a generalization of the Maxwell-Bloch Hamiltonian
which has been used for modelling the interaction of two-level atoms with
radiation \cite{Leclair:1999-1}. Using the polaronic transformation,
$U=\prod_j\exp{\left\{i\beta_j\Phi(j)(\sigma_j^z+2n_j^G-1)/2\right\}}$, (\ref{eq.2})
can be transformed into the double-Kondo Hamiltonian:
\begin{equation}
\label{eq.4}
H_K= \frac{\hbar v}{2}\int_{-L}^0\left(
\hbar^{-2}\Pi^2+(\partial_x\Phi)^2\right)\, \upd x -
\frac{1}{2}\sum_jB^z_j\,\sigma_j^z - \frac{1}{2}\sum_j
E^J_j\,\left(e^{i\beta_j\Phi(j)}\sigma_j^++\ \mathrm{h.c}\,
\right)\,.
\end{equation}
Here $B_j^z=4E^c_j\,(1-2n^G_j)$ where $E^c_j=e^2/2(C^J_j+C^G_j+C_j)$
is the charging energy of the qubit $j$. This model
should not be confused with the
two impurity Kondo model where the two magnetic impurities are
in an infinite environment \cite{Jayaprakash:1981-1}.

In the present physical situation, the total charge stored in the transmission
line and the coupling capacitances is zero (see fig. \ref{fig.1}).
Assuming for simplicity
$\beta_0=\beta_{-L}=\beta$, this fixed charge constraint reads in the Kondo
formulation:
\begin{equation}
\label{eq.8}
\frac{1}{\beta}\int_{-L}^0\hbar^{-1}\Pi(x)\,\upd
x-\frac{1}{2}\,\left(
\sigma_0^z+\sigma_{-L}^z\right)=
n_0^G+n_{-L}^G-1\,.
\end{equation}
Using bosonization and taking care of Klein factors,
the double-Kondo Hamiltonian with the same constraint
can be shown to describe the low-energy physics of
a 1D quantum wire of length $L$ with
resonant levels at its ends (see fig. \ref{fig.2}). The effective
low-energy theory of an isolated finite length Luttinger liquid
\cite{Fabrizio:1995-1}
is characterized by a renormalized Fermi
velocity $v$ and a dimensionless interaction parameter $g$
($g=1$ for the Fermi liquid). This interaction parameter is related to $\beta$
by $\beta^2=\pi/g$.
The Josephson
energies are then proportional to the tunneling
amplitudes between the wire and the resonant levels:
$E^J_j\propto t_j/\sqrt{a}$ where $a$ is a microscopic length scale (UV cutoff).
The magnetic fields $B_j^z$ are
related to the resonant level chemical potentials.
The fixed charge condition (\ref{eq.8}) is
obtained by working in the canonical ensemble: the total number
$\mathcal{N}$ of electrons in the wire
plus the resonant levels is fixed.
In this case, the r.h.s. of (\ref{eq.8}) is equal to $\mathcal{N}+\chi-1$
where $e^{2i\pi\chi}=-e^{i(\theta_{-L}-\theta_0)}$ and $\theta_j$ denotes the phase
relating right to left moving fermions at the wire's
ends: $\psi_R(j)=e^{i\theta_j}\psi_L(j)$.

The one-channel single impurity Kondo model consisting in a spin
1/2 coupled to a free boson on the half line has been shown to be
integrable \cite{Andrei:review,Tsvelik:review}.
Following~\cite{Caux:2002-1} we conjecture that the double-Kondo
problem is integrable for $\beta_0=\beta_{-L}$. This opens the
door to a general study of the thermodynamical properties of the
double-Kondo model using integrable field theory techniques
\cite{Ghoshal:94-1,Fendley:1993-1} as done for the double
sine-Gordon theory \cite{Caux:2003-1}. Such a study would go
beyond the scope of the present letter and we shall therefore
restrict ourselves to the Toulouse point where the problem can be
treated in terms of fermions.

Fermionization of (\ref{eq.4}) is achieved by unfolding the free
non-chiral boson onto a chiral boson $\phi_R(x)$ living on an interval
of length $2L$ and, for $\beta^2=\pi$, introducing the free chiral fermion
$e^{2i\beta\phi_R}$ on $[-L,L]$.
Then, taking $d_{-L}=\sigma_{-L}^+$, $d_{0}=(-1)^{n_{-L}}\sigma_0^+$ and
$\psi_R(x)=(-1)^{n_0+n_{-L}}(2\pi a)^{-1/2}e^{i\sqrt{4\pi}\phi_R(x)}$ ($n_j=(1-\sigma_i^z)/2$)
defines a Jordan Wigner transformation which
turns the boundary spins into fermionic degrees of
freedom anti-commuting with the chiral fermion $\psi_R$.
Because of the fixed charge condition, the fermionic field picks up a
phase between $-L$ and $L$ which depends on the gate
voltages: $\psi_R(x+2L)=-e^{2\pi i\chi}\psi_R(x)$ where
$\chi\equiv n^G_0+n^G_{-L}+1/2\pmod{1}$. The fermionized form of
(\ref{eq.4}) is then:
\begin{equation}
\label{eq.5}
H_F =\int_{-L}^L \; \psi_R^\dagger(x) (-i\hbar v\partial_x) \psi_R(x)\,\upd x -\frac{1}{2}\sum_{j=-L,0}
B_j^z(1-2d_j^\dagger d_j)+\sum_jH^{\partial}_j.
\end{equation}
Boundary interactions $H^{\partial}_j$ are given by:
\begin{eqnarray}
\label{eq.6}
H_{-L}^\partial & = & - \sqrt{\frac{\pi a}{2}} E_{-L}^J e^{i\pi {\cal N}}
\left(e^{-i \pi (\chi-1/2)} d_{-L}^\dagger \psi_R(-L) + h.c. \right)\\
\label{eq.6bis}
H^\partial_{0} & = & -\sqrt{\frac{\pi a}{2}}E_0^J\biggl(d_0^\dagger\,\psi_R(0) + h.c.\biggr)
\end{eqnarray}
where $\mathcal{N}$ denotes the total number of fermions in the system.
Because of the $(-1)^{{\cal N}}$ operator in $H^\partial_{-L}$, we are not yet dealing with free
fermions. The fixed charge contraint (\ref{eq.8}) simply means that the total number of
fermions is fixed as well as the phase $\chi$. Therefore $(-1)^{{\cal N}}$ is a phase that can
be absorbed in a redefinition of $d_{-L}$.
But as we shall see, (\ref{eq.8}) leads to unexpected subtleties in the
computation of the thermodynamical properties of the system as functions of the physical
parameters of the original qubit problem.

Since we are now dealing with free fermions, the ground state energy can
be computed by summing one-particle energies up to the appropriate
Fermi level. These are determined by the stationary wave condition:
\begin{equation}
\label{eq.7}
1+e^{2i(kL-\pi\chi)}\,R_0(k)R_{-L}(k)=0
\end{equation}
where $R_j(k)=(k-k_j-i\xi_j^{-1})/(k-k_j+i\xi_j^{-1})$ denotes the
reflection matrix at boundary $j$. It can be expressed in terms
of the boundary parameters $\xi_j=4\hbar^2v^2/\pi a(E^J_j)^2$ and
$k_j=B_j^z/\hbar v$. As seen from
$R_j(k)$ expression, each qubit influences the modes of the line around
$k_j$ on a momentum scale $\xi_j^{-1}$.
Up to a numerical factor, $\hbar v/\xi_j$ is the energy
scale associated with the Kondo boundary interaction\footnote{The parameter
$g$ in \cite{Leclair:1997-1} corresponds to $\beta^2/2\pi$ in the present letter.
The Toulouse point in \cite{Leclair:1997-1} is at $g=1/2$.}
($\hbar \omega_B$ in \cite{Leclair:1997-1}).
At zero temperature, the finite frequency conductance of a Luttinger
liquid with an impurity as well as the spectral function in dissipative quantum mechanics
are functions of the variable $\omega/\omega_B$ \cite{Lesage:1996-1}. Moreover,
the enveloppe of Friedel oscillations in a Luttinger liquid at
$g=1/2$ is a function of the
variable $\omega_B x/v_F$ \cite{Leclair:1996-1}. In this sense,
$\xi_j$ may be viewed as the Kondo length associated
with each qubit.

We used the technique developed by Chatterjee 
for the 2D Ising model \cite{Chatterjee:1996-1} to obtain
an integral representation for the free energy of the system and
therefore for its ground state energy. The final result is:
\begin{equation}
\label{eq.9}
E_0 = \frac{\hbar v}{\pi}\int_{-\infty}^0\left(L+\frac{\xi_j}{\xi_j^2(k-k_j)^2+1}\right)\;k\,\upd k
-\frac{\hbar v}{2\pi}\int_0^{+\infty}\log{\left(|1+X(ik)|^2\right)}\,\upd k.
\end{equation}
where $X(k)=e^{2i(kL-\pi\chi)}R_0(k)R_{-L}(k)$. The first term, proportional to $L$
is an extensive bulk contribution which does not depend on the
qubit parameters. The second one is a boundary contribution which involves one boundary at
a time. It will therefore not contribute directly to the correlations between the two qubits.
Nevertheless, it contributes to the average population giving back the expected
contribution for a resonant level connected to a semi-infinite Fermi liquid \cite{Furusaki:2002-1}.
Correlations between the qubits are contained in the third term.

\begin{figure}
\onefigure[scale=0.6]{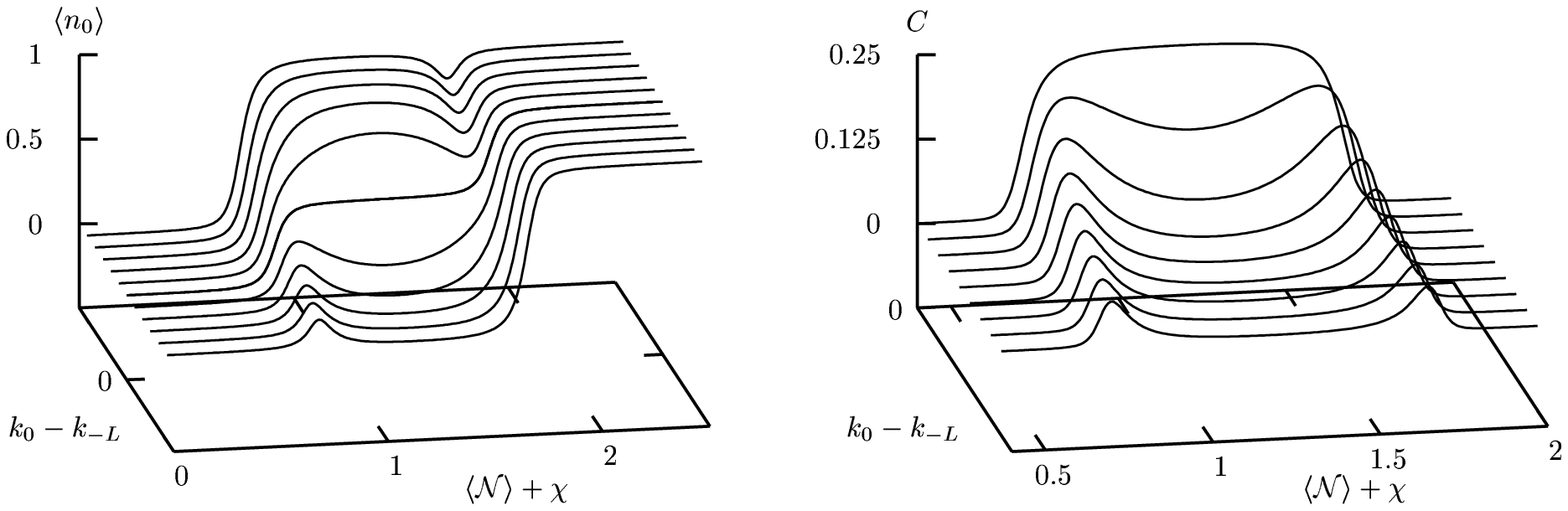}
\caption{Average population $\langle n_0\rangle$ and correlation
$C=\langle n_0\rangle\langle n_{-L}\rangle -\langle n_0\ldotp n_{-L}\rangle$
in the symmetric case $E^J_0=E^J_{-L}$ such that
$\xi_0=\xi_{-L}=200\,L$. For the correlation, only the part with $k_0\geq k_{-L}$ has
been represented since, in this case, $C$ is an even function
of $k_0-k_{-L}$. \label{fig.3}}
\end{figure}

\begin{figure}
\twofigures[scale=0.6]{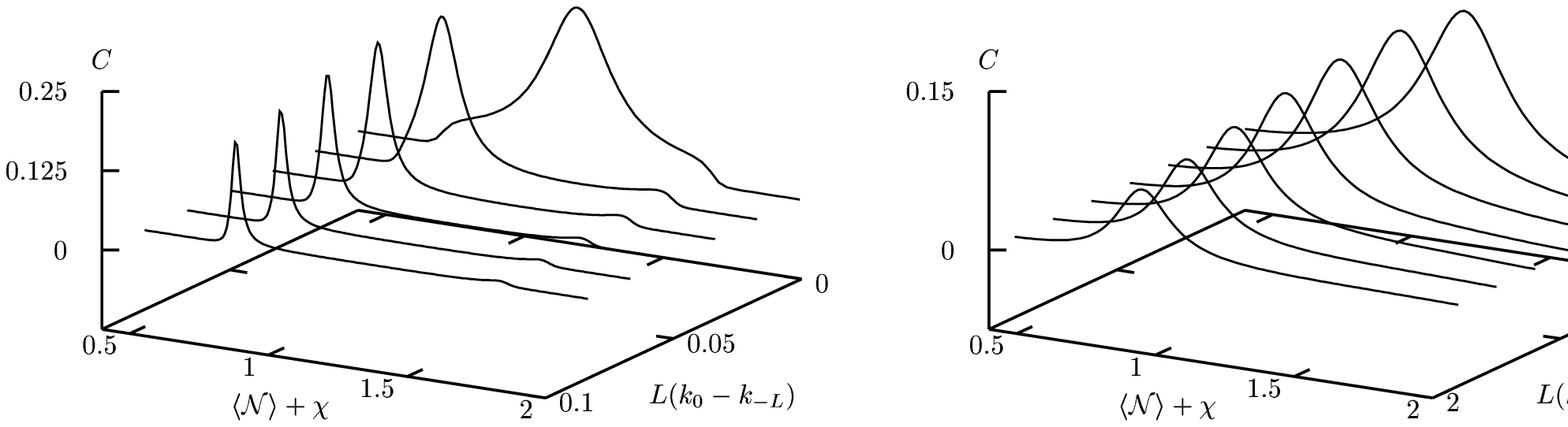}{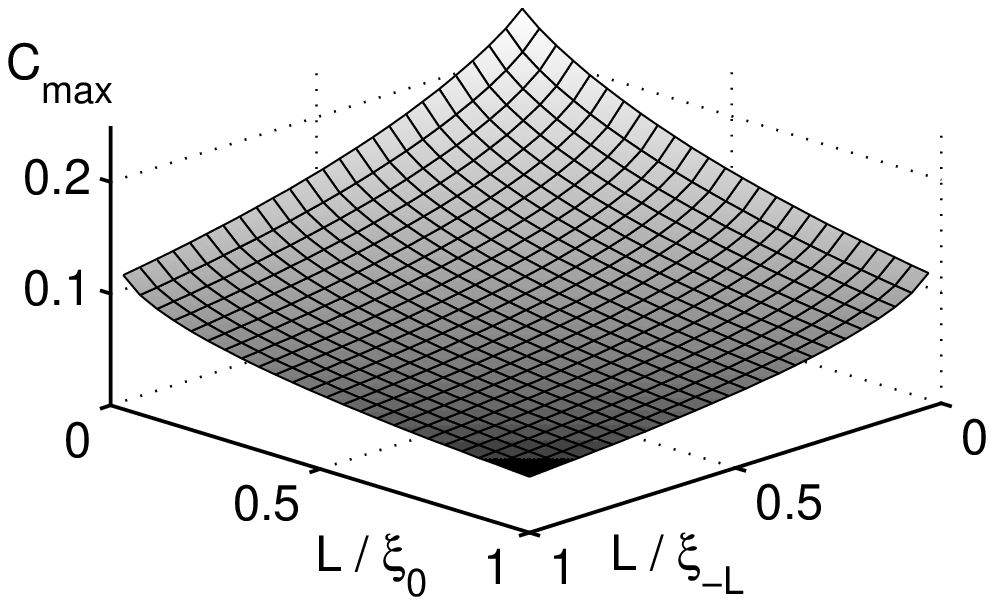}
\caption{The correlation $C$ for $\xi_{-L}/\xi_0=20$, $\xi_0/L=200$.
\label{fig.4}}
\caption{The maximum of $C$ in the $(k_0-k_{-L},\mathcal{N}+\chi)$ plane
as a function of $L/\xi_0$ and $L/\xi_{-L}$.
\label{fig.5}}
\end{figure}

Equation (\ref{eq.9}) should however be taken with care. It corresponds to the energy of a ground state
where only negative eigenstates are filled. But then, each time a one-particle
energy level crosses zero, eq. (\ref{eq.9}) undergoes a discontinuity of its first derivative.
On the other hand, the fixed charge constraint (\ref{eq.8})
tells us that the physical ground state of the system corresponds to a Dirac sea filled
up to a Fermi level that varies with the physical parameters of the
problem without crossing any one-particle energy levels. The physical ground state
energy that incorporates the fixed charge condition should
indeed be an analytical function of the qubit parameters. This is very similar to the analytical
continuation introduced in \cite{Caux:2002-1} for the double sine-Gordon theory.
Using the same method as for the ground state energy (\ref{eq.9})
integral representations for the populations and correlation of the qubits can be obtained and
evaluated numerically, taking into account the fixed charge condition. Results depicted on figures
\ref{fig.3} and \ref{fig.4} are obtained by this procedure.

Let us now discuss the results on correlations between
the qubits induced by the transmission line.
They are expected to depend on the ratios of
$\xi_{-L,0}/L$. Naively, strong
correlations may be expected when at least one of these lengths
becomes comparable or larger than $L$, meaning that both
qubits belong to the same large Kondo cloud. But indeed,
only states whose wave vectors $k$ satisfy $|k-k_j|\xi_j\lesssim
1$ are significantly localized on the site associated with
qubit $j$. Therefore strong correlations only occur when both
Kondo lengths are comparable to $L$. For the same reason, it is also necessary that the two qubits
influence the same modes of the line. This criterion should be seen as a variant of
the exhaustion principle introduced by Nozi\`eres. Finally,
strong correlations occur when both $\xi_j$ are much larger than $L$ and
near degeneracy points $B_0^z\sim B_{-L}^z$.

For $\xi_0\gg L$ and $\xi_{-L}\gg L$
the stationary wave condition (\ref{eq.7}) has two solutions
that tend to $k_0$ and $k_{-L}$ with vanishing Josephson energies.
Correlations appear when one of these special one-particle states is filled. Fig. \ref{fig.3}
displays the average
population of the $x=0$ qubit and the inter-qubit correlation
$C=\langle n_0\rangle\langle n_{-L}\rangle -\langle n_0\ldotp n_{-L}\rangle$
as a function of the difference
$k_0-k_{-L}$ and of $\langle \mathcal{N}\rangle+\chi$ in the regime
$\xi_0=\xi_{-L}\gg L$. Fig. \ref{fig.4} shows the inter-qubit correlation
in terms of the same
variables but for different Kondo lengths illustrating the influence of
asymmetry ($\xi_0\neq \xi_{-L}$): $C$
still reaches its maximal value $1/4$ but in a narrower zone.
The two one-particle states close to $k_{0,-L}$ give
rise to an effective two-level system for the two lowest
energy states of the system. Its effective Hamiltonian
can be computed for $\beta^2< 2\pi$ using perturbation
theory on the double-Kondo Hamiltonian\footnote{IR divergences being cut by the finite
size of the system, only UV divergences should be taken care of.}.
This approach is valid
as long as $LE^J_j/\hbar v \ll \beta^2(L/\pi a)^{\beta^2/2\pi}$.
At degeneracy $B^z_0=B^z_{-L}=B^z$ and for
$0<u<1$, where $u=3/2-(n^G_0+n^G_{-L})+B^zL/\beta^2$,
the effective Hamiltonian in the space generated by
$|\!\uparrow\downarrow\rangle \otimes |\beta (n^G_0+n^G_{-L}-1),\{0\}\rangle$ and
$|\!\downarrow\uparrow\rangle\otimes|\beta (n^G_0+n^G_{-L}-1),\{0\}\rangle$
is of the form $\Delta(X\sigma^z+\sigma^x)$ where
\begin{equation}
\Delta = \left(\frac{\pi a}{L}\right)^{\frac{\beta^2}{\pi}}
\frac{E_{-L}^J E_0^J}{4\pi\hbar v/L}
\frac{\Gamma(\frac{\beta^2}{\pi}u)\Gamma(\frac{\beta^2}{\pi}(1-u))}{\Gamma(\frac{\beta^2}{\pi})}
\ \mathrm{and}\ X=\left(
\left|\frac{E_0^J}{E_{-L}^J}\right|-\left|\frac{E_{-L}^J}{E_0^J}\right|\right)\frac{\sin{(\beta^2(u-1/2))}}
{2\sin{(\beta^2/2)}}\,.
\end{equation}
The correlation is then equal to
$C=1/4(1+X^2)$ and the energy separation
between the ground state and the first excited state is given by $2\Delta\sqrt{1+X^2}$.
For $E^J_0=E^J_{-L}$, the two-qubit entanglement is maximal and the
ground state is given by $\frac{1}{\sqrt 2}
(|\!\uparrow\downarrow\rangle + |\!\downarrow\uparrow\rangle)\otimes
|\beta (n^G_0+n^G_{-L}-1),\{0\}\rangle$. Note that
nevertheless entanglement is lost when going back to the original qubit
problem because of the orthogonality effect induced by the polaronic
transformation: qubit states $|\downarrow\uparrow\rangle$ and
$|\uparrow\downarrow\rangle$ get correlated to different states of the transmission line
whose scalar product goes as $(a/L)^{\beta^2/\pi}$.
As expected, entanglement can appear in the original qubit problem for small
$\beta^2$: for $E^c_j,\,E^J_j\ll \hbar v/L$
integration over the modes of the line gives an effective static coupling of
the form $(v\beta^2/4L)\,\sigma_0^z\sigma_{-L}^z$. This is the regime usually
considered for quantum computation.

When one of the $\xi_j$ is much smaller than $L$, the corresponding qubit renormalizes onto a
conformally invariant boundary condition. This regime is reached
in the limit of very large Josephson energy and corresponds to the
strong coupling limit of the Kondo problem. The corresponding
quantum impurity is totally screened and therefore is not seen
by the other one. Then, when either
$\xi_0\ll L$ or $\xi_{-L}\ll L$, the correlation $C$ is expected to vanish
(Kondo regime).
This is confirmed by our exact results but we
can also study the crossover where one of
the $\xi_j$ is much larger than $L$ and the other varies. For $L\ll \xi_{-L}$ and
$\xi_0\sim L$,
our analysis for $\beta^2=\pi$ predicts a maximum absolute value of correlation between
the qubits decaying as $1/4(1+L/\xi_0)$. Fig. \ref{fig.5} displays
the maximal $C$ as a function of $L/\xi_0$ and $L/\xi_{-L}$.

To summarize, we have described a common approach to the problem of two qubits coupled via a finite
length quantum transmission line and to the problem of two resonant levels
at the ends of a finite length quantum wire. These two problems can be mapped
to the double-Kondo model with a fixed charge constraint. At its Toulouse point, we have solved
the model exactly enabling an explicit analysis of the
correlation between the quantum impurities (qubits or resonant levels) through their
common environment. The emergence of correlations is related to the overlap
of the influence zones (Kondo clouds) of each quantum impurity
in reciprocal space. We stress that the same techniques
can be extended to obtain results on dynamical correlation
fonctions. This gives a direct access to the absorbtion spectrum
of the system when excited by a microwave radiation sent on one of the
available gates. The correlations between the charge of each qubit and the charge density
along the line can also be obtained exactly as well as the Friedel oscillations
induced by the resonant levels at the ends in the quantum wire problem. These correlations
provide a direct insight into the Kondo cloud in real space.

Based on our study at the Toulouse point and on
the perturbative approach at $\beta^2<2\pi$, our
conclusions on the role of Kondo lengths on inter-qubit
correlations should hold for $\beta^2<2\pi$. This remains to be fully confirmed by analytical
computations in the non perturbative regime away from $\beta^2=\pi$. From an experimental
point of view, the quantum wire design is suitable for exploring the region around
the Toulouse point. Since transmission line impedances are in the current status of
technology bounded by a few hundred Ohms, the qubit design is unfortunately limited
to $\beta \lesssim 0.3$ but both schemes provide complementary ways to explore the same
physics.

\acknowledgments
P. Degiovanni would would like to acknowledge Boston University for hospitality while
this work was completed and C. Chamon for a careful reading of the manuscript.

\end{document}